\documentclass[runningheads]{llncs}

\usepackage[T1]{fontenc}
\usepackage{amsmath, amssymb, amsfonts}
\usepackage{graphicx}
\usepackage{booktabs, array}
\usepackage{verbatim}
\usepackage{overpic}
\usepackage{xspace, xcolor}
\usepackage{chngcntr}
\counterwithout{table}{section}
\usepackage{xspace}

\newcommand{\name}{\emph{VesselSDF}\xspace}

\def\ie{\emph{i.e.,}\xspace} 

\begin{document}

\title{\name: Distance Field Priors for Vascular Network Reconstruction}

\author{Salvatore Esposito\inst{1} \and
Daniel Rebain\inst{2} \and
Arno Onken\inst{1} \and
Changjian Li\inst{1} \and \\
Oisin Mac Aodha\inst{1}}

\authorrunning{S.~Esposito et al.}

\institute{
School of Informatics, University of Edinburgh, United Kingdom\and
Department of Computer Science, University of British Columbia, Canada\\
\email{sesposit@ed.ac.uk}
}

\maketitle

\begin{abstract}
Accurate segmentation of vascular networks from sparse CT scan slices remains a significant challenge in medical imaging, particularly due to the thin, branching nature of vessels and the inherent sparsity between imaging planes. Existing deep learning approaches, based on binary voxel classification, often struggle with structural continuity and geometric fidelity. To address this challenge, we present \name, a novel framework that leverages signed distance fields (SDFs) for robust vessel reconstruction. Our method reformulates vessel segmentation as a continuous SDF regression problem, where each point in the volume is represented by its signed distance to the nearest vessel surface. This continuous representation inherently captures the smooth, tubular geometry of blood vessels and their branching patterns. We obtain accurate vessel reconstructions while eliminating common SDF artifacts such as floating segments thanks to our adaptive Gaussian regularizer which ensures smoothness in regions far from vessel surfaces while producing precise geometry near the surface boundaries. Our experimental results demonstrate that \name significantly outperforms existing methods and preserves vessel geometry and connectivity, enabling more reliable vascular analysis in clinical settings.

\end{abstract}

\section{Introduction}
Accurate reconstruction of vascular networks from medical imaging data remains a fundamental challenge in clinical diagnostics and surgical planning. Blood vessels, with their intricate branching patterns and varying diameters, play a crucial role in numerous pathologies, from coronary artery disease to liver tumor assessment \cite{Xu2025VSNet}. However, their thin, tree-like structures present unique challenges for reconstruction, particularly when working with sparse CT scan slices. The clinical implications are significant -- precise vessel reconstruction enables better surgical navigation, improves understanding of blood flow dynamics, and aids in early detection of vascular abnormalities \cite{Tetteh2020}.

Traditional segmentation and reconstruction approaches and recent deep learning methods often struggle with the inherent sparsity between imaging planes, leading to discontinuities and loss of critical geometric features~\cite{Zhao2019}. While deep learning has shown promise in medical image segmentation \cite{Ronneberger2015}, current approaches face two critical challenges: (i) maintaining structural coherence and (ii) generalizing beyond the training data. The first challenge manifests in regions where vessels branch or significantly change direction between consecutive slices, often resulting in fragmented or anatomically implausible reconstructions~\cite{Zhou2020}. The second challenge stems from the inherent scarcity of medical imaging datasets, especially those with annotated thin structures~\cite{Elsayed2024Vessel}, leading to models that memorize specific vessel configurations rather than learning generalizable geometric principles.

Signed distance field (SDF) based representations offer a promising solution to these challenges due to their inherent ability to represent smooth, continuous surfaces~\cite{Park2019}. By encoding geometry through distance fields, SDFs naturally capture thin structures and maintain consistent spatial relationships \cite{Alblas2022OffGrid}.
Inspired by this, we present \name, a novel approach that leverages these geometric principles while addressing common SDF-based reconstruction artifacts.
Our approach combines a specialized two-stage neural SDF encoder-decoder architecture with a novel Gaussian regularization technique that eliminates floating artifacts by adaptively enforcing smoothness based on distance from the vessel surface. To address the generalization challenge, we reformulate vessel reconstruction as continuous geometric regression rather than discrete voxel classification, allowing the model to learn underlying shape principles that transfer across different vessel configurations and anatomical variations. Our distance-weighted regularization further enhances generalization by encoding universal geometric priors related to vessel continuity rather than memorizing specific patterns. This is complemented by an efficient SDF refinement strategy that ensures robust reconstruction even in challenging cases with significant inter-slice gaps.

In this work, we make the following contributions:
(i) We introduce \name, a novel neural SDF architecture for vessel segmentation and reconstruction that enables accurate reconstruction of thin vascular structures from sparse CT slices.
(ii) We propose an adaptive Gaussian regularization technique for SDF learning that applies distance-weighted smoothness constraints, effectively reducing floating geometric artifacts while preserving fine vessel details.
(iii) We demonstrate superior reconstruction performance on challenging clinical vessel data  containing thin vessels and complex branching patterns.

\section{Related Work}
\noindent{\bf Vasculature Segmentation.} Deep learning approaches have largely superseded traditional methods in medical image segmentation. The foundational U-Net~\cite{Ronneberger2015} architecture established key principles through its encoder-decoder structure with skip connections. However, vessel segmentation remains challenging due to the presence of thin branching structures and high inter-subject variation~\cite{Wittmann2024VesselFM}. The segmentation of fine vascular structures presents unique difficulties, as minor vessels are frequently undetected or fragmented in standard frameworks. This challenge stems from their limited spatial extent and complex topology, creating a fundamental tension between preserving connectivity and maintaining accurate boundaries~\cite{Dima2023b}. Recent innovations have tackled this problem through specialized architectures and loss formulations, such as  topology-aware loss functions to penalize disconnected vessels~\cite{clDiceOriginal,clDice}. Concurrently, architectures have evolved to better capture multi-scale vessel features, with transformer-based models~\cite{Zhou2023nnFormer,Shaker2024UNETRpp} leveraging self-attention mechanisms to model global vessel connectivity while preserving convolutional inductive bias~\cite{Chen2021TransUNet}. Hybrid approaches that combine convolution with transformers have proven effective for CT segmentation~\cite{Li2022}. Structure-aware networks explicitly incorporate vascular morphology through vessel-growing decoders in multi-task frameworks~\cite{Xu2025VSNet}. Implicit neural representations have also been utilized~\cite{Alblas2022OffGrid} to produce continuous, topologically consistent vessel surfaces, and domain-adaptive models have been shown to generalize across imaging modalities~\cite{Wittmann2024VesselFM}. These developments demonstrate a shift toward specialized techniques that address the fundamental challenges of thin structure segmentation.

\noindent{\bf Sparse Slice Segmentation.}
A significant challenge in medical image segmentation is handling sparse slice data~\cite{Wu2022SliceImputation}, which occurs due to radiation dose reduction in CT or time constraints in MRI. This limited through-plane resolution creates discontinuities in vessel connectivity, particularly at bifurcations or directional changes~\cite{Yagis023}. While deep learning approaches have shown promise in general medical segmentation, accurately reconstructing continuous vessel structures under sparse sampling remains difficult, especially with complex vascular topologies. Recent approaches have explored 3D shape constraints~\cite{Alblas2022OffGrid,Tetteh2020}, but the problem persists for fine vessel structures.

\noindent{\bf Geometric Priors and Consistency.}
Recent efforts have integrated anatomical knowledge and geometric constraints to improve segmentation accuracy, especially when data are sparse or noisy. Topology-preserving approaches explicitly penalize disconnected or anatomically implausible predictions~\cite{clDiceOriginal}, improving continuity in thin vascular segments~\cite{clDice}. For instance, by incorporating synthetic angiogram training data and cross-hair filters to better handle vessel continuity in large volumetric scans~\cite{Tetteh2020}.
Beyond segmentation, advances in neural rendering have shown strong potential for capturing thin vessel structures through continuous SDFs~\cite{Bogensperger2023,Ma2021,walker2025crosssdf} and neural radiance fields~\cite{Maas2023}. In contrast, our \name approach poses vessel segmentation as a continuous SDF regression problem with an adaptive Gaussian regularizer, which simultaneously preserves fine vessel geometry near surface boundaries while ensuring smoothness in distant regions, eliminating floating geometric artifacts.

\begin{figure}[t]
\centering
\begin{overpic}[width=1.0\linewidth]{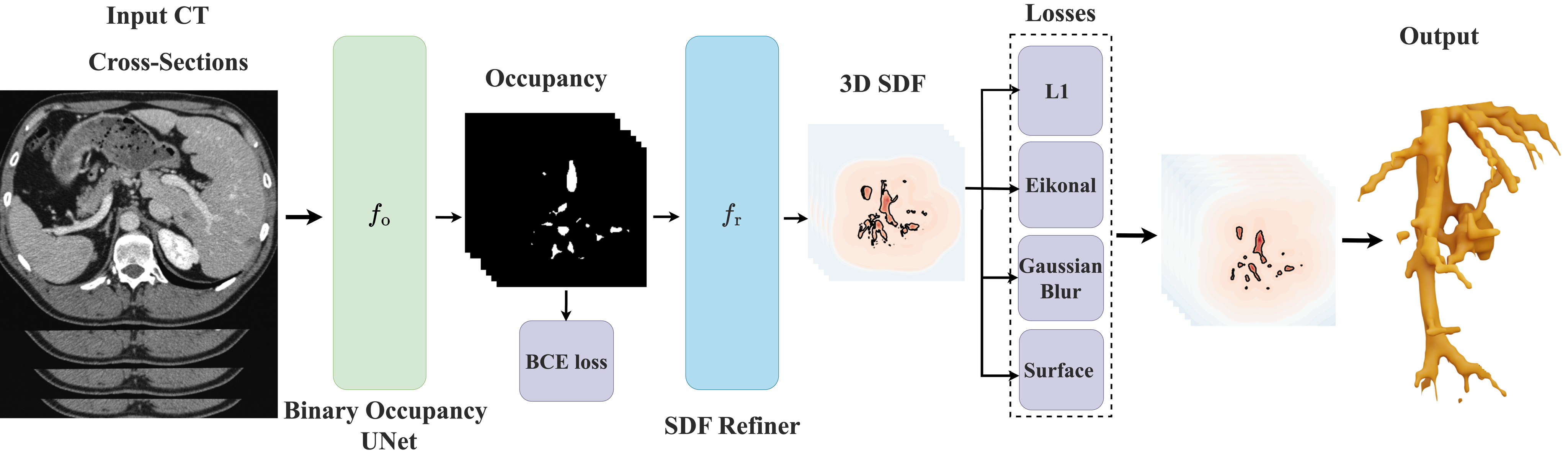}
\end{overpic}
\caption{{\bf Overview of \name}, our two-stage approach for vessel segmentation and reconstruction from CT scans. In the first stage, a 3D U-Net predicts a binary occupancy map. The second stage refines this occupancy into a signed distance field (SDF) using an additional 3D U-Net,
guided by geometric regularization terms.
The output 3D SDF, converted into a mesh, contains high-quality reconstructed vessels.
}
\label{fig:teaser-fig}
\end{figure}

\section{Method}

\subsection{Problem Statement}

Traditional binary vessel segmentation approaches face three critical limitations. First, the discrete nature of voxel-based representations results in jagged surface artifacts, which are particularly pronounced in thin vessels where the surface-to-volume ratio is high. Second, the substantial difference between in-plane resolution \(\Delta x, \Delta y\) and slice thickness \(\Delta z\) creates anisotropic distortions that fragment vessel structures, especially at branching points. Third, while signed distance fields (SDFs) offer a promising direction for continuous surface representation, existing SDF-based methods often generate floating artifacts, \ie disconnected surface fragments that degrade reconstruction quality.

We address these challenges via \name, a two-stage framework that explicitly separates vessel segmentation from geometric reconstruction (see Fig.~\ref{fig:teaser-fig} for an overview). \name takes a volumetric CT scan \(\mathcal{V} \in \mathbb{R}^{D \times H \times W}\) as input, where \(D\) is the number of axial slices, and \((H, W)\) are the in-plane dimensions. As output, it produces a continuous signed distance field (SDF) $f_{\mathrm{SDF}}(\mathbf{x}; \theta_r)$, where \(\mathbf{x}\) is a 3D spatial coordinate and \(\theta_{r}\) are the model parameters,
which implicitly defines the vessel surface through its zero-level set:
\begin{equation}
\label{eq:zero-level-set}
S := \bigl\{\mathbf{x} \in \mathbb{R}^3 \,\mid\, f_{\mathrm{SDF}}(\mathbf{x}; \theta_r) = 0\bigr\}.
\end{equation}

\subsection{Binary Occupancy (Stage 1)}

The first stage of \name focuses on binary vessel segmentation via an occupancy prediction function $f_{o}(\mathbf{x}; \theta_{o})$.
By focusing initially on the segmentation task, we establish a reliable starting point for subsequent geometric refinement.  This occupancy predictor is implemented via a 3D encoder–decoder CNN-based U-Net architecture that captures multi-scale vessel features and outputs a voxelwise vessel occupancy probability.

Accurate segmentation of thin, branching vessels requires multi-scale feature representation.  To achieve this, we integrate 3D attention gates~\cite{oktay2018attention} such that  at each level \(\ell\) of the encoder–decoder structure, we combine gating feature maps \(\mathbf{g}_{\ell}\) with skip-connection feature maps \(\mathbf{h}_{\ell}\) through a learned attention mechanism:
\begin{equation}
\label{eq:attention-gates}
\boldsymbol{\alpha}_{\ell}
=
\psi\!\Bigl(\mathbf{W}_{g}\,\mathbf{g}_{\ell} \;+\; \mathbf{W}_{h}\,\mathbf{h}_{\ell}\Bigr),
\end{equation}
where \(\psi(\cdot)\) is a learned attention function that highlights salient vessel features, and \(\mathbf{W}_{g}\) and \(\mathbf{W}_{h}\) are trainable weight matrices.  The resulting attention weights \(\boldsymbol{\alpha}_{\ell}\) modulate the skip-connection features, ensuring focus on critical vessel regions and preserving fine vascular details that might otherwise be overlooked.

\subsection{SDF Refinement (Stage 2)}

The second stage transforms the binary occupancy into a correctly scaled SDF:
\begin{equation}
\label{eq:sdf-refinement}
f_{\mathrm{SDF}}(\mathbf{x}; \theta_r)
\;=\;
f_{r}\bigl(\operatorname{detach}\bigl(f_{o}(\mathbf{x}; \theta_{o})\bigr);\,\theta_{r}\bigr),
\end{equation}
where \(f_{r}(\cdot;\theta_{r})\) is our SDF refiner network.
The network regresses the signed distance to the surface. The gradient detachment ensures that SDF‐specific distance constraints do not interfere with the initial segmentation task. This refinement is key to achieving smooth, accurate vessel reconstructions.  Rather than treating each voxel independently, the SDF representation inherently couples neighboring predictions through distance‐field properties.  Each level operates at a different spatial resolution through successive downsampling and upsampling operations, allowing the network to simultaneously capture fine vessel details at high resolutions while modeling long-range dependencies and global vessel connectivity at lower resolutions. This hierarchical design allows the network to capture both fine details and global connectivity.

\subsection{Optimization}
\name's training loss combines supervised learning and geometric constraints to ensure  continuous vessel reconstructions that are both accurate and topologically coherent, even for thin, branching vessels:
\begin{equation}
\label{eq:total-loss}
\mathcal{L}
=
\lambda_s\,\mathcal{L}_{\mathrm{sdf}}
\;+\;
\lambda_o\,\mathcal{L}_{\mathrm{occ}}
\;+\;
\lambda_e\,\mathcal{L}_{\mathrm{eik}}
\;+\;
\lambda_g\,\mathcal{L}_{\mathrm{gauss}}
\;+\;
\lambda_r\,\mathcal{L}_{\mathrm{sur}}.
\end{equation}

\noindent{\bf Supervised Terms.}
We use both SDF and occupancy supervision:
\begin{equation}
\label{eq:sdf-loss}
\mathcal{L}_{\mathrm{sdf}}
=
\mathbb{E}_{\mathbf{x} \in \Omega}
\Bigl\lvert
  f_{\mathrm{SDF}}(\mathbf{x}) - f_{\mathrm{SDF}}^{*}(\mathbf{x})
\Bigr\rvert,
\end{equation}
\begin{equation}
\label{eq:occ-loss}
\mathcal{L}_{\mathrm{occ}}
=
-\mathbb{E}_{\mathbf{x} \in \Omega}
\Bigl[
  y\,\log \bigl(f_{o}(\mathbf{x})\bigr)
  +
  (1 - y)\,\log \bigl(1 - f_{o}(\mathbf{x})\bigr)
\Bigr],
\end{equation}
where \(f_{\mathrm{SDF}}(\mathbf{x})\) is our SDF prediction, \(f_{\mathrm{SDF}}^{*}(\mathbf{x})\) is the ground-truth SDF, \(y \in \{0,1\}\) is the binary vessel label, and \(\Omega \subset \mathbb{R}^3\) is the domain of 3D spatial coordinates in our training volume. Note that \(\Omega\) includes the axial dimension as well, so these terms are computed fully in 3D (\ie across neighboring slices), thereby ensuring consistent supervision throughout the volumetric scan.

\noindent{\bf Eikonal Regularization.}
To encourage smooth distance transitions, we enforce near‐unit gradients \cite{implicit_2020}:
\begin{equation}
\label{eq:eikonal-loss}
\mathcal{L}_{\mathrm{eik}}
=
\mathbb{E}_{\mathbf{x} \in \Omega}
\Bigl[
  \bigl(\bigl(\partial_x f_{\mathrm{SDF}}(\mathbf{x})\bigr)^2
      + \bigl(\partial_y f_{\mathrm{SDF}}(\mathbf{x})\bigr)^2
      + \bigl(\gamma\,\partial_z f_{\mathrm{SDF}}(\mathbf{x})\bigr)^2\bigr)
  \;-\;
  1
\Bigr]^2,
\end{equation}
where \(\gamma = \tfrac{\Delta z}{\Delta x}\) accounts for anisotropic voxel spacing along the axial dimension. By enforcing this constraint, the model more closely approximates the signed distance property, preventing large deviations in gradient magnitudes that can lead to geometric artifacts.

\noindent{\bf Distance-weighted Gaussian Regularization.}
While the Eikonal term enforces local smoothness in the gradient field, high-frequency noise can still persist in regions far from vessel boundaries. To further suppress such artifacts, we introduce a \emph{distance-weighted Gaussian} regularizer:
\begin{equation}
\label{eq:gaussloss}
\mathcal{L}_{\mathrm{gauss}}
=
\mathbb{E}_{\mathbf{x} \in \Omega}
\Bigl\lvert
  f_{\mathrm{SDF}}(\mathbf{x})
\Bigr\rvert
\cdot
\Bigl\|
  f_{\mathrm{SDF}}(\mathbf{x})
  \;-\;
  \mathcal{G}_{\sigma}\!\bigl(f_{\mathrm{SDF}}(\mathbf{x})\bigr)
\Bigr\|_{2}^{2}.
\end{equation}
Here, \(\mathcal{G}_{\sigma}\) denotes a 3D Gaussian blur operator with scalar standard deviation \(\sigma > 0\). This term smoothens the SDF \emph{more aggressively} in regions where \(\lvert f_{\mathrm{SDF}}(\mathbf{x})\rvert\) is large (\ie\ far from the vessel surface) and thus more prone to noisy fluctuations, while preserving fine details near the boundary (\ie where \(\lvert f_{\mathrm{SDF}}(\mathbf{x})\rvert \approx 0\)). This distance‐adaptive mechanism is key to achieving globally smooth reconstructions without over-smoothing critical vessel edges.

\noindent{\bf Surface Regularization.}
Finally, to suppress spurious or ``floating” vessel components, we add:
\begin{equation}
\label{eq:surface-loss}
\mathcal{L}_{\mathrm{sur}}
=
\mathbb{E}_{\mathbf{x} \in \Omega}
\exp\!\Bigl(-\beta\,\bigl\lvert f_{\mathrm{SDF}}(\mathbf{x})\bigr\rvert\Bigr),
\end{equation}
where \(\beta>0\) is a hyperparameter that controls how strongly near-zero SDF values are penalized when there is no strong evidence of an actual surface. Larger \(\beta\) more aggressively suppresses weak, noisy boundaries.

\section{Experiments}
\noindent{\bf Implementation.}
\name is supervised on 3D SDF slices.
For the occupancy stage, we employ a 3D U-Net with attention gates that produces voxel-wise occupancy logits.
A second lighter 3D U-Net with two encoder–decoder levels refines these detached occupancies into the final SDF.
Both stages are trained jointly for 100 epochs without data augmentation to preserve accurate SDF values using the Adam optimizer with a learning rate of $5\times10^{-4}$.
Training is performed on whole-volume inputs of size $512\times512\times16$ with a batch size of 16, and marching cubes is applied at a resolution of 512 to extract the output meshes. The loss weights are set to
$\lambda_s=0.1,\ \lambda_o=0.01,\ \lambda_e=0.01,\ \lambda_g=0.1,$ and $\lambda_r=0.1$.

\noindent{\bf Datasets.}
We evaluate \name on two public hepatic vessel segmentation datasets. The Hepatic Vessels dataset (Medical Segmentation Decathlon - Task 08)~\cite{Antonelli_2022} contains 303 hepatic veins CT scans with semi-automated vessel annotations obtained via level-set based region growing from expert-placed seed points.
The ground truth includes major hepatic venous structures (portal and hepatic veins) and hepatic arteries visible in portal-phase CT.
The IRCADb dataset~\cite{soler20103d} contains 20 contrast-enhanced abdominal CT scans with fully manual  segmentations of liver vascular structures, focusing primarily on hepatic and portal vein branches.
We compute training SDFs from the binary ground truth.

\noindent{\bf Baselines.}
We compare to three state-of-the-art volumetric segmentation architectures which perform binary voxel classification. 3D-UNet~\cite{cciccek20163d}, which is a standard encoder-decoder architecture for medical image segmentation. 3D SA-UNet~\cite{guo20233d} extends this architecture with spatial attention modules that adaptively weight feature responses based on their relevance to vessel structures, particularly beneficial for capturing thin vasculature.
nnU-Net~\cite{isensee2021nnu} which contains two U-Nets, a low resolution and a higher resolution one.

\noindent{\bf Evaluation.}
Our evaluation uses five metrics to assess 3D vessel reconstruction: Dice Score measures volumetric overlap; Volume IoU assesses 3D segmentation precision; Jaccard Distance (JD) quantifies topological similarity; and Chamfer Distance (CD) and Hausdorff Distance (HD) evaluate geometric accuracy through average and maximum surface distances.

\begin{table}[t]
    \begin{center}
    \caption{{\bf Quantitative Results on the Hepatic Vessels and IRCADb datasets.}
    Comparison of vessel reconstruction performance using different baselines. We report volume metrics (Dice Coefficient, Intersection over Union (IoU), and Jaccard similarity (JD)) and surface metrics (Chamfer distance (CD) $\times 100$ and Hausdorff Distance (HD)).}
    \label{tab:combined_results}
    \setlength{\tabcolsep}{12pt}
    \resizebox{0.80\columnwidth}{!}{%
    \begin{tabular}{l|ccccc}
        \hline
        \multicolumn{6}{c}{\bf Hepatic Vessels Dataset} \\
        \hline
        Model & Dice↑ & IoU↑ & JD↑ & CD↓ & HD↓ \\
        \hline
        \name (Ours)       & \textbf{0.72} & \textbf{0.59} & \textbf{0.48} & \textbf{0.68} & \textbf{4.1} \\
        nnU-Net~\cite{isensee2021nnu} & 0.69 & 0.56 & 0.45 & 0.82 & 4.9 \\
        3D SA-UNet~\cite{guo20233d}      & 0.64 & 0.51 & 0.38 & 1.3  & 5.2 \\
        3D-UNet~\cite{cciccek20163d}        & 0.53 & 0.44 & 0.31 & 1.6  & 5.6 \\
        \hline
        \multicolumn{6}{c}{\bf IRCADb Dataset (Portal and Hepatic Veins)} \\
        \hline
        \name (Ours)       & \textbf{0.86} & \textbf{0.82} & \textbf{0.75} & \textbf{0.60} & \textbf{3.5} \\
        nnU-Net~\cite{isensee2021nnu}  & \textbf{0.86} & \textbf{0.82} & \textbf{0.75} & 0.75 & 4.2 \\
        3D SA-UNet~\cite{guo20233d}      & 0.85  & 0.80 & 0.72 & 0.80 & 4.5 \\
        3D-UNet~\cite{cciccek20163d}        & 0.84  & 0.79 & 0.72 & 0.90 & 5.0 \\
        \hline
    \end{tabular}
    }
    \end{center}
\end{table}

\subsection{Results}
\begin{figure}[t]
\centering
\includegraphics[width=1.0\linewidth]{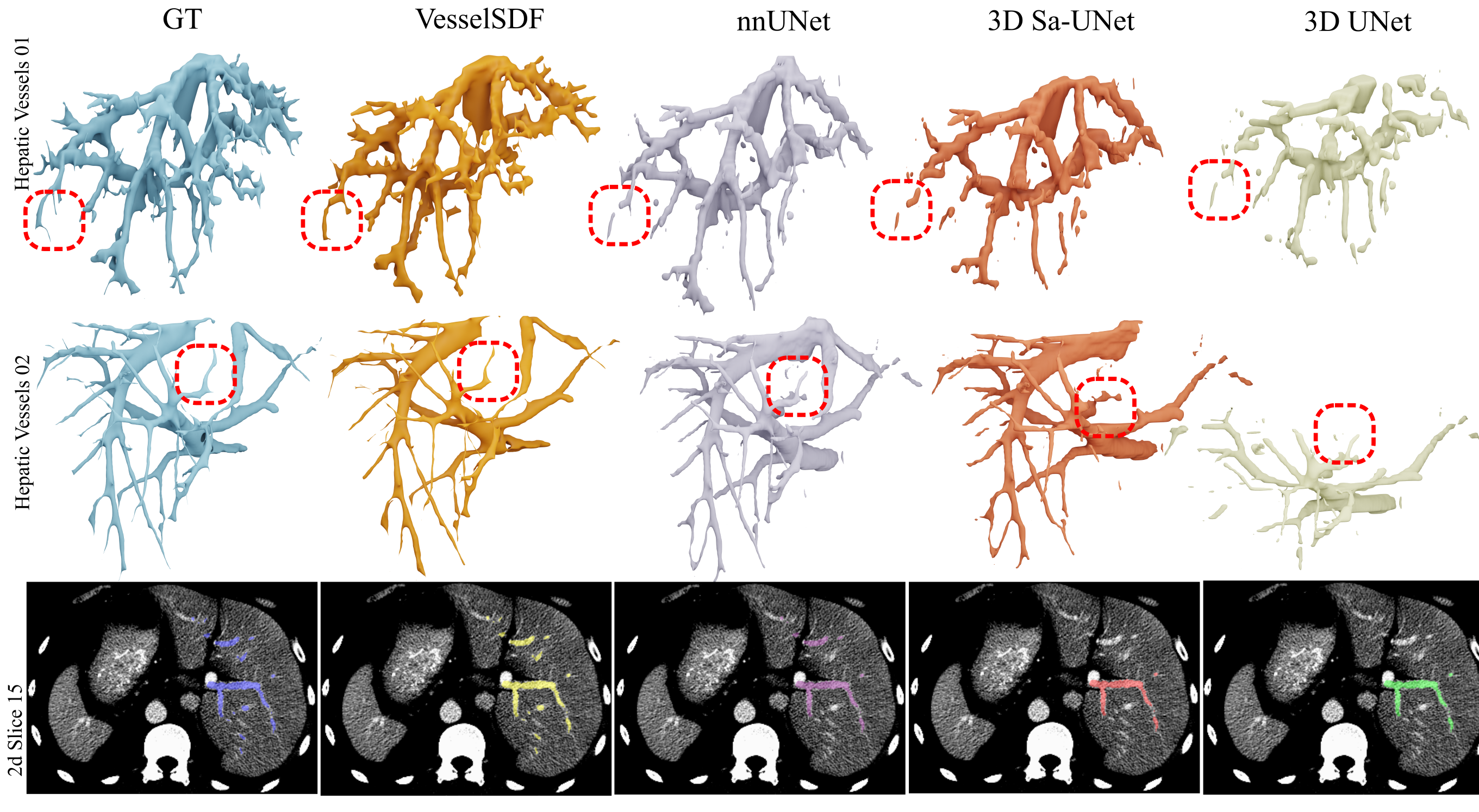}
\caption{{\bf Qualitative  3D reconstruction results on the Hepatic Vessels dataset.} The bottom row displays 2D slices highlighting the segmentation results.
}
\label{fig:results_hepatic}
\end{figure}

We present quantitative results in Table~\ref{tab:combined_results} and qualitative results in Fig.~\ref{fig:results_hepatic} where we demonstrate \name's superior  vessel and portal vein reconstructions on the Hepatic Vessels~\cite{Antonelli_2022} and IRCADb~\cite{soler20103d} datasets.
Our approach preserves thin vessels and complex branching structures more effectively than the binary voxel classification baselines (nnU-Net, 3D SA-UNet, and 3D-UNet), yielding more complete and anatomically coherent reconstructions.
Quantitatively, we outperform all methods on the challenging Hepatic Vessels dataset and obtain comparable performance according to volume-based metrics on IRCADb, where the numerical results are in general similar across all method, but we significantly outperform the baselines according to the surface-based metrics.

\begin{table}[h]
    \begin{center}
    \caption{{\bf Ablations on the Hepatic Vessels dataset.}}
    \label{tab:ablation_results}
    \resizebox{0.72\columnwidth}{!}{%
    \begin{tabular}{l|ccccc}
        \hline
        Model & Dice↑ & IoU↑ & JD↓ & CD↓ & HD↓ \\
        \hline
        \name                       & \textbf{0.72} & \textbf{0.59} & \textbf{0.48} & \textbf{0.68} & \textbf{4.1} \\
        \name w/o SDF refinement    & 0.69          & 0.57          & 0.52          & 0.70          & 4.4          \\
        \name w/o Binary Occupancy  & 0.65         & 0.55          & 0.56          & 0.75          & 4.6          \\
        \name w/o Gaussian Loss     & \textbf{0.72}          & \textbf{0.59}          & \textbf{0.48}          & 0.70          & 4.3          \\
        \hline
    \end{tabular}
    }
    \end{center}
\end{table}

\noindent{\bf Ablations.} Table~\ref{tab:ablation_results} quantifies the contribution of each component in \name.
\textit{w/o SDF refinement:} we remove the second SDF refiner, using only binary occupancy prediction without distance field computation or geometric regularizers.
\textit{w/o Binary Occupancy:} we directly predict the complete SDF (surface and isolines), bypassing our two-stage approach to test whether separating vessel detection from geometric refinement is beneficial.
\textit{w/o Gaussian Loss:} we remove the adaptive regularization (Eq.~\eqref{eq:gaussloss}), which results in similar Dice scores but introduces surface artifacts.
The full \name achieves best  performance across all metrics, with SDF refinement particularly enhancing vessel continuity as shown by the improved reconstruction metrics.

\section{Conclusion}
We presented \name a new approach for thin vessel segmentation and reconstruction.
Via a carefully designed two-stage refinement architecture with geometric regularization, \name addresses many of the geometric artifacts that can get introduced by a naive application of SDFs to this task.
We demonstrate superior quantitative and qualitative 3D reconstructions compared to baselines on challenging sparse CT slice data containing hepatic vessels, where our results exhibit fewer issues such as floating geometry and disconnected structures.

\clearpage
\begin{credits}
\subsubsection{\ackname} SE was funded by the UKRI CDT in Biomedical AI.

\subsubsection{\discintname}
The authors have no competing interests to declare that are relevant to the content of this article.
\end{credits}

\bibliographystyle{splncs04}
\bibliography{main}

\end{document}